\documentclass{optica-article}

\journal{opticajournal} 

\articletype{Research Article}

\usepackage{fancyhdr}
\usepackage{lineno}

\usepackage[ 
            ]{todonotes}
            
\usepackage{placeins}

\usepackage{braket}
\usepackage{commath} 
\usepackage[print-unity-mantissa = false, range-units = single]{siunitx}

\graphicspath{{./figures/}}

\def\equationautorefname~#1\null{Eq.~(#1)\null}

\begin{document}

\title{Detecting clear-air turbulence via beam broadening in a Rayleigh-scattering lidar system}

\author{ Christopher Miller\authormark{1,*,+}, Daniel Lum\authormark{1,+}, Brandon Rodenburg\authormark{1,+},  Michael Stenner\authormark{1}, Anthony DiCarlo\authormark{1}, Bradford Snios\authormark{1}, and Paul D. Williams\authormark{2}}

\address{\authormark{1}The MITRE Corporation, 7525 Colshire Dr. McLean, VA 22102, USA}
\address{\authormark{2}University of Reading, Brian Hoskins Building, Whiteknights Rd, Earley Gate, Reading, RG6 6ET, UK}
\authormark{+}These authors contributed equally.
\authormark{*}Corresponding Author: \email{cwmiller@mitre.org} 


\begin{abstract*} 
The volume of clear-air turbulence (CAT) in the atmosphere at flight cruising altitudes is increasing rapidly, posing a growing problem for civil aviation and resulting in reduced confidence in aviation safety. There are limited remote detection capabilities for CAT, since clear air produces no measurable radar return. Lidar has been proposed as a viable detection methodology, and several systems have been demonstrated. However, these systems have to date demonstrated limited detection ranges of less than 15~km. In this work, we propose a novel lidar-based CAT detection methodology that uses Rayleigh scattering and relies on a differential detector measurement to quantify beam spread and thereby estimate the eddy dissipation rate (EDR), which is the international aircraft-independent metric for quantifying aviation turbulence strength. Additionally, we present experimental results demonstrating the validity of the optical efficiency model used in the detection simulations. We show that, under modest assumptions, a size, weight, and power (SWAP) constrained system that implements this method can detect moderate CAT at ranges in excess of 30~km, equating to two minutes of flight time for typical commercial aviation cruising speeds, which represents a substantial range improvement over prior approaches. This is an important advance because---for the first time---it potentially allows the cabin to be secured before the turbulence is encountered, reducing the injury risk to passengers and flight attendants.

\end{abstract*}

\section{Introduction}
Turbulence poses a growing threat to the aviation industry, affecting safety, operational efficiency, and airline economics. Turbulence is already the leading cause of injuries to passengers and cabin crew, leading to higher medical costs, compensation claims, and flight diversions \cite{sharman2012}. Operationally, airlines face higher fuel consumption and maintenance costs as aircraft deviate from optimal flight paths to avoid turbulent regions. Moreover, turbulence-related delays reduce schedule reliability and passenger confidence. Addressing turbulence risk is therefore critical for sustaining aviation safety and resilience.

As an exacerbating factor, turbulence is becoming more severe and prevalent. The main phenomenon generating clear-air turbulence (CAT), vertical wind shear, has increased by 15\% at flight cruising altitudes over the North Atlantic since modern satellite observations began in 1979 \cite{lee2019}. 
For this reason, the volume of severe CAT at flight cruising altitudes over the North Atlantic and USA has increased by 55\% and 41\%, respectively, since 1979 \cite{prosser2023}. Future projections indicate large increases in CAT in the coming decades \cite{smith-williams}.

Although increases in near-cloud turbulence associated with convective cells in cloudy air are concerning, this turbulence can, in principle, be avoided by the use of airborne radar. By contrast, for turbulence in clear air, there does not currently exist any mature remote detection system to give aircraft sufficient warning to change altitudes or secure the cabin. Clear air is invisible to radar, so there is no radar return from CAT. Often, the only warning of oncoming CAT is a report from an aircraft ahead that has already encountered the turbulence. The current lack of remote detection methods for CAT, coupled with the long-term rise in the prevalence of CAT, makes it imperative to develop on-board technologies to remotely sense turbulence in clear air.

One possible remote detection method for CAT relies on lidar \cite{zhao2023a}. Clear air scatters a small but non-negligible proportion of light ranging from ultra-violet (UV) to near-infrared (NIR) wavelengths $(280~\text{nm}$--$2~\mu \text{m})$. As such, lidar systems can be used to probe parcels of air that would not produce a measurable return from radar.

Prior lidar systems have detected CAT by making a measurement of the air velocity, using either Doppler or Rayleigh (incoherent) methods. A Doppler lidar measures the Doppler frequency shift in the returned signal reflected by aerosols and thereby infers the velocity of the air. A collaboration between Boeing and the Japan Aerospace Exploration Agency (JAXA) conducted tests using a 1.55~$\mu$m IR Doppler lidar and found that the CAT detection range was 17.5~km averaged over all altitudes but dropped to under 10~km at 40,000~feet \cite{matayoshi2018}. This drop-off occurs because Doppler lidar systems have limited range at higher altitudes where the concentration of aerosols is lower. In contrast, a Rayleigh lidar can operate in aerosol-free air and measures fluctuations in the back-scatter coefficient resulting from fluctuations in air density associated with turbulence. A European Union consortium project called DELICAT (DEmonstration of LIdar based Clear Air Turbulence detection) conducted tests using a 355~nm UV Rayleigh lidar and found that the detection range for moderate CAT was 10~km \cite{vrancken2016}. Unfortunately, the deviations of the volumetric scattering coefficient from turbulence are relatively small, and so a large return signal is required to extract the signal from the noise. While UV lidar can produce strong returns because air is a very strong scatterer of UV, this same advantage reduces the maximum range of these devices. Also, UV Rayleigh lidar can fail when aerosol scattering contaminates the return signal. Visible (532~nm) Rayleigh lidar has been proposed as an alternative, exploiting an iodine absorption filter to prevent this contamination. Theoretical analyses indicate a possible detection range of 7.5--11.7~km for moderate-to-severe CAT using this method \cite{zhao2023b}, but in-flight tests to verify these theoretical predictions have not yet been carried out.

In this work, we propose an alternative lidar detection method to quantify turbulence levels. Unlike Doppler lidar, our method does not rely on aerosols and remains effective in the cleaner upper atmosphere. Unlike previous efforts that employed Rayleigh-scattering lidar, our detection method does not attempt to directly measure fluctuations in density. Instead, we propose to quantify turbulence using a beam-broadening Rayleigh-scattering lidar system, by measuring the excess beam spread associated with increased turbulence. Beam-broadening is a well-known phenomenon in astronomy and optical communications; a wavefront transmitted through a turbulent channel will deform as it propagates. To first order, this deformation results in a broadening of the beam at range, relative to a beam that traversed a clear channel. By binning photons returning to the system into inner and outer rings, we can measure the change in beam width associated with transmission through turbulence. This phenomenon of optical distortion does not depend on the presence of aerosols and results in a stronger signal than measurement of density fluctuations based on quantifying the change in volumetric backscattering coefficient.

We present theoretical analysis and simulation results that demonstrate that a system with limited SWAP is capable of measuring moderate-to-severe turbulence at ranges in excess of 30~km, which is much greater than the detection ranges reported in previous work. We also provide criteria for optimal wavelength selection and demonstrate that the proposed detection architecture is near optimal among all architectures that function by measuring beam spread. We derive a detection model that maps collected counts in both detectors to a posterior distribution of the eddy dissipation rate (EDR).

\section{Optical Turbulence Modeling}
In this section we describe the theoretical model for CAT that is used to develop the detection theory. First, we present our model for the optical efficiency, the ratio of collected to emitted photons, for Rayleigh lidar systems along with an analysis of optimal wavelength selection as a function of desired range. Following that we present analysis for the radial spread of a lidar pulse in the presence of turbulence.

\subsection{Optical Efficiency Model}\label{sec:Link_Budget}
The optical efficiency, the fraction of emitted lidar power that returns to the receiver, is a key design parameter for any lidar-based turbulence detection system. Without adequate return photons, no amount of signal processing can hope to infer any information about the atmospheric channel. For this discussion we assume a mono-static configuration of lidar emitter and receiver apertures. However, a bistatic configuration can be modeled with small modifications to the equations presented below. 

We assume that a pulsed lidar system with pulse energy $J_0 = \tau P_0$, where $\tau$ is the pulse duration and $P_0$ is the peak power, is emitted into the atmosphere. As the beam propagates, it is attenuated by absorption and scattering effects so that the pulse energy $J$ incident on a plane perpendicular to the direction of travel at range $z$ is given by
\begin{equation}\label{attenuation}
    J(z) = J_0\exp(-\alpha z),
\end{equation}
where $\alpha [\textrm{m}^{-1}]$ is the total attenuation (absorption + scattering) per unit range. We assume that the air has a molecular scattering cross section $\sigma = \sigma(\lambda) [\textrm{m}^2]$ and a volumetric scattering coefficient $\beta [\textrm{m}^{-1}]= \sigma n$, where $n [\mathrm{m^{-3}}]$ is the number density of air molecules per cubic meter. The infinitesimal pulse energy scattered over a path length $\dif z$ is given by
\begin{equation}\label{scattered-energy}
    dJ_{scat} = \beta J \dif z.
\end{equation}
We assume operation in the upper atmosphere and thus we neglect the contributions of aerosols and attribute all scattering to Rayleigh scattering. We assume that the receiver is a circular aperture of radius $R$. The energy scattered in the direction of the receiver can be found by integrating the unit-normalized Rayleigh scattering phase function over the solid angle subtended by the circular aperture. The unit-normalized Rayleigh scattering phase function is given by
\begin{equation}\label{phase-fn}
\psi(\theta) = \frac{3}{16 \pi}(1 + \cos^2 \theta),    
\end{equation}
and we define the function $\Psi(z; R)$ to be the integral of \eqref{phase-fn} over the aperture viewed from a point at $z$,
\begin{align}\label{phase-integral}
\Psi(z;R) &= \int_0^{2\pi}\int_0^\omega(z) \psi(\theta) \sin(\theta) \dif\theta\ \dif\phi = \frac{1}{8}(-c_\omega^3 - 3c_\omega + 4),
\end{align}
where $\omega$ is the half angle of the cone subtended by the receiver aperture when viewed from distance $z$ and $c_\omega \equiv \cos \omega = z/\sqrt{z^2 + R^2}$.
The differential energy scattered back towards the receiver, given by \eqref{scattered-energy} and \eqref{phase-integral}, is
\begin{equation}
    dJ_{backscatter} = \beta \Psi(z) J(z) \dif z.
\end{equation}
If we consider integration over a small range cell of length $\delta z$ then the total energy scattered from that cell in the direction of the receiver is given by 
\begin{equation}\label{backscattered-energy}
    J_{BS}(z) = \beta \Psi(z) J(z) \delta z.
\end{equation}
Combining \eqref{backscattered-energy} with \eqref{attenuation} accounting for two-way propagation losses yields the energy returned to the receiver for a single pulse as
\begin{equation}\label{eq:total-returned-energy}
J_{rcv}(z) = J_0\exp(-2\alpha z) \beta \Psi(z) \delta z.
\end{equation}

\begin{figure}[htbp]
    \centering
    \includegraphics[width=0.8\linewidth]{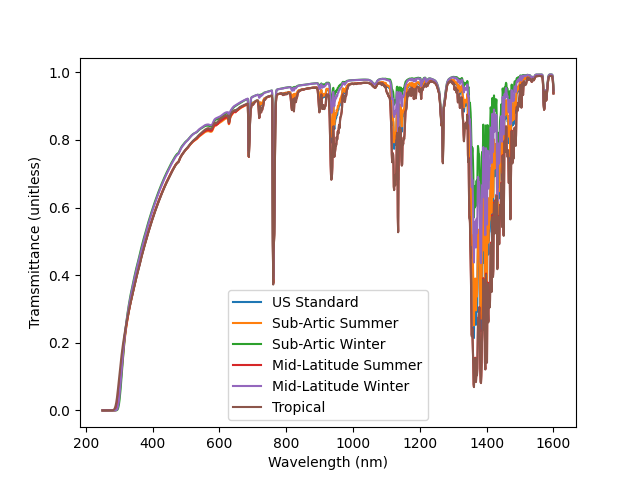}
    \caption{MODTRAN-modeled transmission over 30~km path.}
    \label{fig:trans}
\end{figure}

\begin{figure}[htbp]
    \centering
    \includegraphics[width=0.8\linewidth]{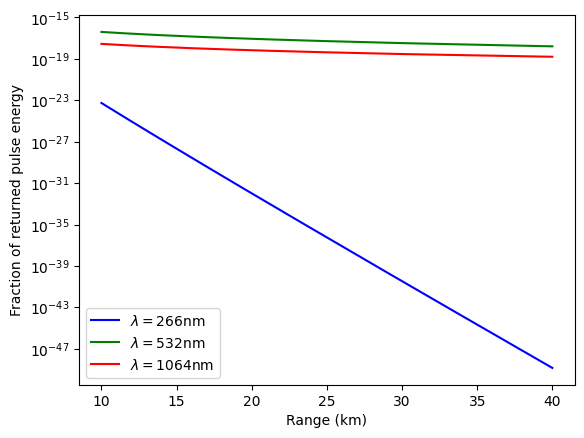}
    \caption{Optical efficiency as a function of wavelength and range.}
    \label{fig:link}
\end{figure}

In order to fix the optical efficiency in a realistic scenario we need to specify notional values for the atmospheric parameters. The scattering cross-section of various gasses at a wavelength of 532~nm can be found in \cite{sneep2005direct}. Since Rayleigh scattering scales with the inverse of the wavelength to the 4\textsuperscript{th} power, we estimate the value of the scattering cross section for other wavelengths by scaling the value at 532~nm by a factor of $\delta \lambda^{-4}$. The values at 266~nm, 532~nm, and 1064~nm are given by
\begin{equation}
\begin{aligned}
    \sigma_{266} &= 8.16\times 10^{-30}\ [\textrm{m}^2],\\
    \sigma_{532} &= 5.1\times 10^{-31}\ [\textrm{m}^2],\ \text{and}\\
    \sigma_{1064} &= 3.19\times 10^{-32}\ [\textrm{m}^2].
\end{aligned}
\end{equation}
The volumetric scattering coefficient is given by $\beta = \sigma n$, where $n$ is the number density of air molecules per unit volume. We use MODTRAN with the US Standard atmospheric profile to obtain a nominal value for $c = 16.1 [\textrm{mol}/ \textrm{m}^3]$ at an altitude of 9~km by taking the MODTRAN-derived temperature and pressure and employing the ideal gas law. The attenuation coefficient $\alpha$ is also obtained using MODTRAN. The transmissivity of the atmosphere at 9~km altitude over a 30~km path is shown in \autoref{fig:trans} for the six MODTRAN model atmospheres. Using the US Standard atmospheric profile we obtain the attenuation coefficients for various wavelengths of interest as
\begin{equation}
\begin{aligned}
\alpha_{266} &= 0.93\ [\rm{km}^{-1}],
\\
\alpha_{532} &= 0.007\ [\rm{km}^{-1}],\ \text{and}
\\
\alpha_{1064} &= 0.002\ [\rm{km}^{-1}].
\end{aligned}
\end{equation}
With these parameters we show return energy as a function of range for three systems with wavelengths in the ultraviolet (UV), green, and near-infrared (NIR) wavelengths in \autoref{fig:link}.

\subsection{Optimal Wavelength as a Function of Desired Range}
There is an interesting symmetry to this sensing task in that the return mechanism, backscattering, is also the main contributor to losses. That is, if the selected wavelength avoids a absorption band, the attenuation coefficient is mainly due to volumetric scattering. Since scattering scales with $1/\lambda^4$ this suggests that there is an optimal wavelength for a given desired range where the scattering contributes as much as possible from the volume at range without being so severe that no energy makes it to that volume.  Below we formalize this idea and derive an expression for the wavelength that optimizes the signal for a given desired range.

\begin{figure}[htbp]
    \centering
    \includegraphics[width=0.8\linewidth]{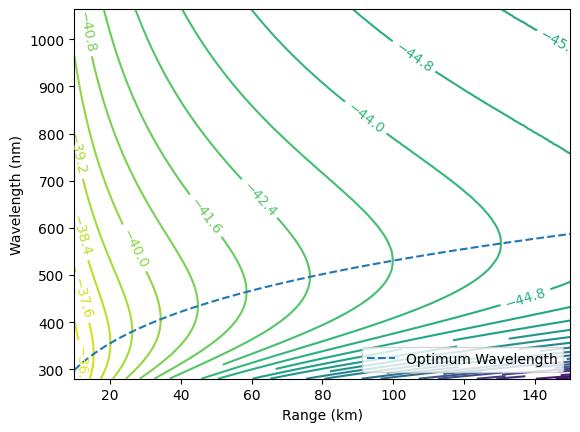}
    \caption{Contours of log-returned energy fraction with optimal wavelength as function of range.}
    \label{fig:opt_wl}
\end{figure}

We assume that absorption is a negligible contributor to the attenuation term in \eqref{attenuation} and rewrite that equation as
\begin{equation}\label{scattering-only-link}
J_{rcv}(z) \approx J_0 \exp(-2\beta z) \beta \Psi(z)\delta z.
\end{equation}
If we assume $\beta(\lambda) = \beta_0 \lambda_0^4 / \lambda^4$ and drop terms that don't depend on $z$ we obtain
\begin{equation}\label{scattering-fn-lambda}
J_{rcv}(\lambda ; z) \approx \exp\left(-2 \beta_0 \frac{\lambda_0^4}{\lambda^4}z \right) \beta_0 \frac{\lambda_0^4}{\lambda^4} \Psi(z).
\end{equation}
Taking the derivative with respect to $\lambda$ and solving for the critical point gives
\begin{equation}
    \lambda_{opt}(z) = \lambda_0 (2\beta_0 z)^{1/4}.
\end{equation}
A plot of the optimal wavelength as a function of desired detection range is shown in \autoref{fig:opt_wl}.

This result does neglect a few key effects. First and somewhat obviously, the optimum is predicated on the scattering being the dominant attenuation term and thus does not hold if the described optimum is in an atmospheric absorption band. Second it also neglects the fact that background radiance will degrade the signal-to-noise ratio and is also wavelength-dependent. Ideally a band could be identified near the optimum wavelength where the background radiance is also small. We used MODTRAN to simulate radiance curves for several solar sensor geometries and failed to identify such a band for daytime operation. Because of this, the optimality criterion above is still a useful heuristic but the advantages of one wavelength over another may be somewhat overstated. The third effect applies specifically to our proposed detection method. The beam wavelength also governs diffraction: since we intend to image the backscattered beam, the receiver field of view must comfortably encompass the diffracted spot. Since the background noise scales with the field of view we generally prefer shorter wavelengths to reduce the effect of background. 

\subsection{Effect of Turbulence on Optical pulses}
The primary mechanism by which turbulence affects optical propagation is the generation of random fluctuations in the refractive index $n(\mathbf r)$. These fluctuations are quantified by the refractive-index structure function 
  \begin{equation}
    \Braket{\left[n(\mathbf{r_1}) - n(\mathbf{r_2})\right]^2} = C_n^2\delta r^{2/3},
\end{equation}
where $\delta r = \|\mathbf{r_1 - r_2}\|$ and $C_n^2$ is the index of refraction structure parameter which characterizes the relative strength of the optical turbulence and will range from roughly \qtyrange{e-18}{e-15}{m^{-2/3}}\cite{Andrews2023, Rodenburg2014, Rodenburg2015}.

\begin{figure}[htpb]
    \centering
    \includegraphics[width=0.8\textwidth]{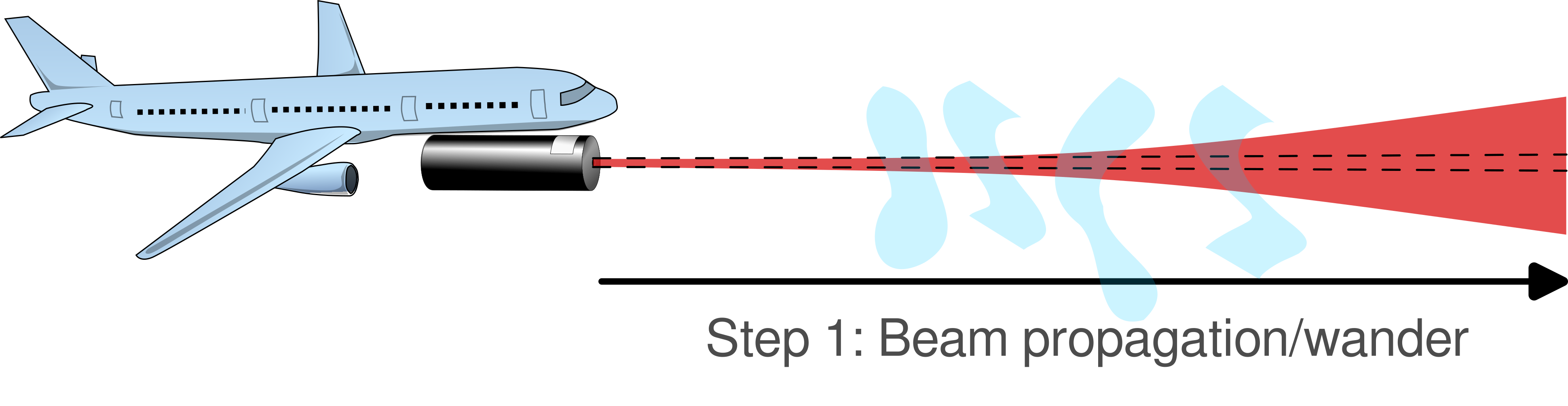}
    \includegraphics[width=0.8\textwidth]{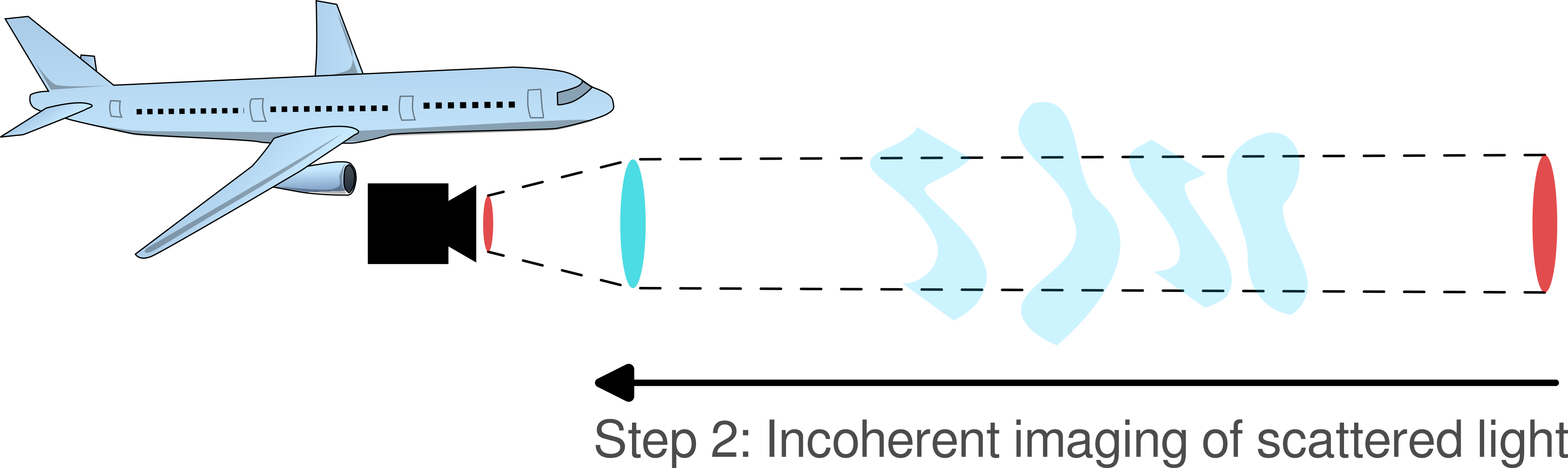}
    \caption{Top: Beam cross section grows over propagation distance due to diffraction as well as turbulence induced expansion from beam wander and scintillation. Bottom: Backscattered light from the beam produces an incoherent source that is imaged through the turbulence. The resulting image is affected by diffraction as well as the turbulence.}
    \label{fig:TwoStepModel}
\end{figure}

The random fluctuations affect both the outward‑propagating beam and the backscattered light, as illustrated in \autoref{fig:TwoStepModel}. The effect of outward beam propagation will include beam wander and scintillation due to randomly accumulated spatial phase front. We assume our initial beam is Gaussian with field profile
\begin{equation}
    U \propto \exp(-r^2/W_0^2)\exp(-ikr^2/2F_0),
\end{equation}
where $W_0$ is the launch waist, and $F_0$ is the phase-front curvature chosen to minimize the beam waist at the target range. In the absence of turbulence, the intensity of the field after a propagation of $z$ from the receiver can be written as the Gaussian function, $I(\mathbf{r}) \propto \exp(-2r^2/W(z)^2).$ The time-averaged effect on the beam of turbulence can be expressed as\cite{Andrews2023}
\begin{equation}
    \Braket{I(r,z)} \propto \exp \left(-2r^2/W_{LT}^2\right),
    \label{eq:TurbSpreadBeam}
\end{equation}
which has a modified beam size of 
\begin{equation}
    W_{LT} = W\left(1 + 1.33\sigma_R^2\Lambda^{5/6}\right)^{3/5},
    \label{eq:WLT}
\end{equation}
where  $\sigma_R^2 = 2.25 k^{7/6}\int_0^L C_n^2(z)z^{5/6}\dif z$ is the Rytov variance and $\Lambda = 2z/kW^2$ is the nondimensional Gaussian beam parameter. If the turbulence is independent of distance, the refractive index structure parameter is constant, $C^2_n(z) = C_n^2$ and the Rytov variance becomes $\sigma_R^2 = 1.23C_n^2k^{7/6}L^{11/6}.$

At every region of space, light is scattered back towards the receiver as described in \autoref{sec:Link_Budget}. A collection optic can be used to focus light, forming an image at the focus. In the absence of turbulence this image will be limited by the collection optics and diffraction. However, the turbulent atmosphere creates a spatially random optical wavefront that increases the point spread function of the formed image. This can be quantified by the optical transfer function (OTF) of the combined system and atmosphere $H = H_\text{atm}H_\text{sys}.$ If the system is diffraction-limited then the turbulence-free OTF is given by 
\begin{equation}
H_\text{sys}(f) = 
    \begin{cases}
        \frac{2}{\pi}\left[\arccos\left(\frac{|f|}{2f_0}\right) - \frac{|f|}{2f_0}\sqrt{1-(f/2f_0)^2}\right] &\text{for $f\le 2f_0$}\\
        0 &\text{otherwise}
    \end{cases},
\label{eq:H_sys}
\end{equation}
where $f$ is the spatial frequency, $f_0 = D/(F\lambda)$ is the optical cut-off frequency in terms of the aperture diameter $D$, and $F$ is the focal length of the system. We can also approximate this as a Gaussian blur with OTF $H_\text{sys}\approx\exp\left[-(f/f_0)^2\right]$.  The turbulence OTF is\cite{TurbulenceOTF2017}
\begin{equation}
    H_\text{atm}
        = \exp\left[-3.44\left(\frac{D}{r_0}\frac{|f|}{f_0}\right)^{5/3}\right] 
        \approx  \exp\left[-\left(3.44^{3/5}\frac{D}{r_0}\frac{f}{f_0}\right)^2\right], 
    \label{eq:H_atm}
\end{equation}
where Fried's parameter is given by $r_0 = \left(0.16k^2\int_0^L C_n^2(z)\dif z\right)^{-3/5}$. If the turbulence is independent of distance, the refractive index structure parameter is constant, $C^2_n(z) = C_n^2$ and Fried's parameter becomes $r_0 = \left(0.16k^2C_n^2L\right)^{-3/5}$.
 
\begin{figure}[htpb]
    \centering
    \includegraphics[width=0.85\textwidth]{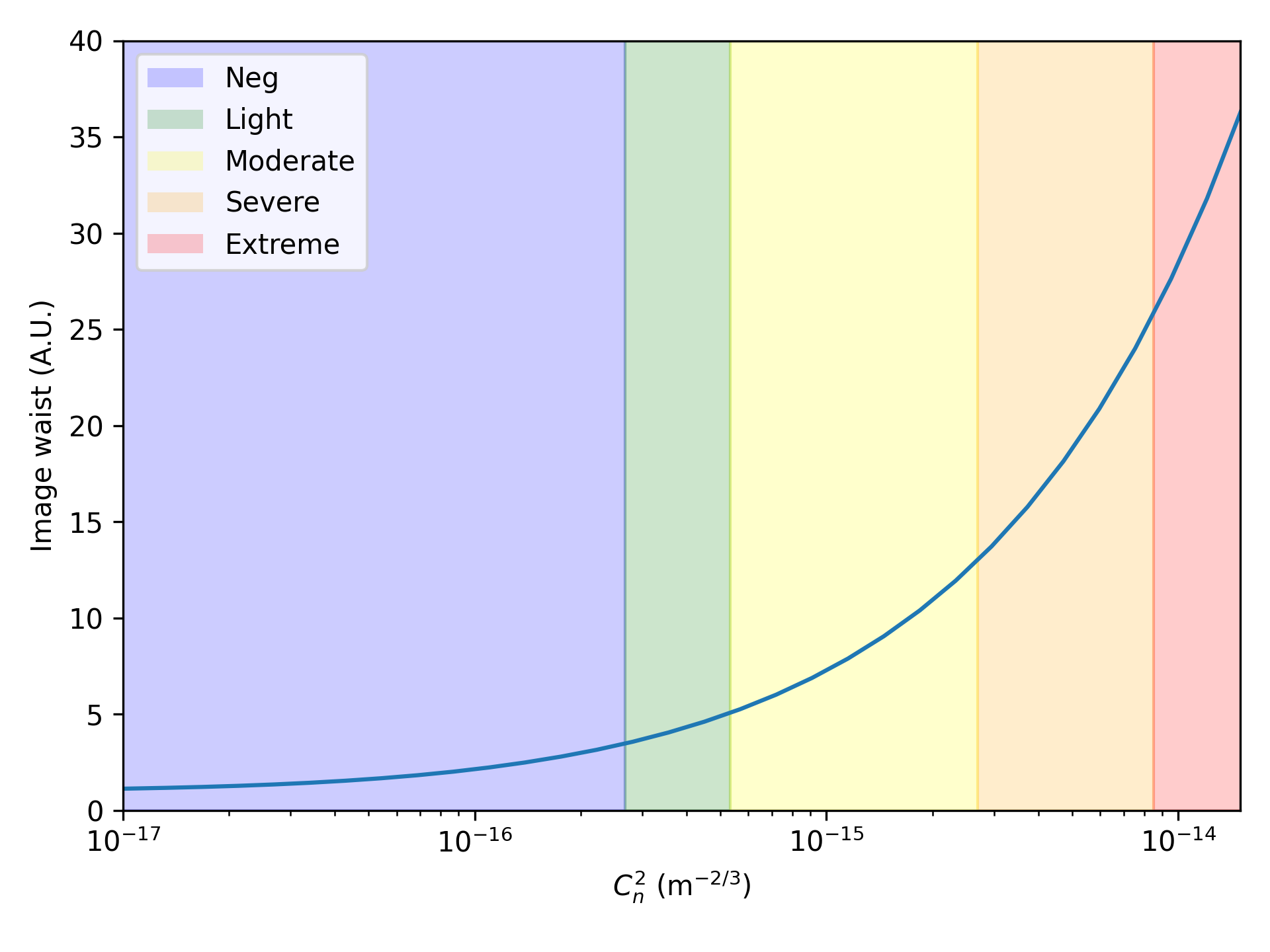}
    \caption{Plot of image spot size formed by an image of the beam at a range of 30~km vs $C_n^2$. Assumes a focused beam at $\lambda = \qty{532}{nm}$ with width $W_0 = \qty{3.4}{cm}$ and receiver aperture of $D=\qty{10}{cm}$. Shaded regions represent (from left to right) negligible, light, moderate, severe, and extreme turbulence regions respectively.}
    \label{fig:WimagevsCn2}
\end{figure}

The geometric image (ignoring diffraction and turbulent induced aberrations) is a Gaussian of width $F/z W_{LT} \approx F\theta_{LT}$, where $\theta_{LT}$ is the angular width of the source as seen from the receiver. The image is the convolution with the Point Spread Function (PSF), which is the Fourier transform of the OTF.  If we use the approximate Gaussian forms of \autoref{eq:H_sys} and \eqref{eq:H_atm}, we can express the average image spot analytically as $I_\text{image} \propto \exp(-r^2/W_i^2),$ where the beam waist is given by
\begin{equation}
W_i^2 = \frac{1}{2}\left(\frac{DW_{LT}}{f_0\lambda z}\right)^2
        +  4.4\left(\frac{D}{\pi r_0 f_0}\right)^2
        + \frac{1}{(f_0\pi)^2}.
\label{eq:W_image}
\end{equation}
The three terms in \autoref{eq:W_image} correspond to the geometric image of \autoref{eq:TurbSpreadBeam}, and the blurring due to $H_\text{atm}$ and $H_\text{sys}$ respectively. A plot of $W_i$ from backscattered light from $L=\qty{30}{km}$ is shown in \autoref{fig:WimagevsCn2} assuming constant turbulence along the path characterized by $C_n^2(z) = C_n^2$. When simulating potential hardware designs that implement our approach, we also apply a magnification factor to magnify the image of the beam up to a useable size. This factor does not affect the theoretical development of the approach but has implications for ultimate design of the hardware. We list the value used in our simulations in \autoref{sec:sim_params}.

\section{Proposed Detection Methodology}
\subsection{Information Bounds for a Differential Detector}
To estimate the optimal detection configuration for measuring turbulence we consider the amount of information contained within the collected back-scattered light. Suppose we are interested in determining the effective refractive index structure over a given path distance we are interrogating. We define a unit-less parameter $\theta$ such that
$$C_n^2 \equiv \exp(\theta)\,\text{m}^{-2/3}.$$
Every photon that falls upon the image plane gives information about $\theta$, as the image waist is a function of $C_n^2$ and thus $\theta$. We can quantify the amount of information per detected photon contained within the image plane intensity by using the Fisher Information $\mathcal I(\theta)$ defined as
\begin{equation}
\mathcal I(\theta) = \left\langle\left(\dpd{}{\theta}\log p(\mathbf{r}|\theta)\right)^2\right\rangle,
      \label{eq:FI_definition}
\end{equation}
where $p(\mathbf r|\theta)$ is the conditional probability density function of measuring a photon at location $\mathbf{r}$ conditioned on $\theta$ which is given by the normalized image intensity function 
\begin{equation}
    p(\mathbf r|\theta) = I_\text{image} = \frac{\exp\left(-r^2/W_i^2(\theta)\right)}{\pi W_i^2(\theta)}.
    \label{eq:p(r|theta)}
\end{equation}

By substituting \autoref{eq:p(r|theta)} into \autoref{eq:FI_definition} we can compute the Fisher Information in the image,
\begin{equation}
\mathcal I_\text{image}(\theta)
    = \int \left(\dpd{}{\theta}\log p(\mathbf{r}|\theta)\right)^2p(\mathbf{r}|\theta)\dif\mathbf{r},
    = \frac{4}{W_i^2}\left(\dpd{W_i}{\theta}\right)^2,
\label{eq:I_image}
\end{equation}
where we have used the identity
\begin{equation}
    \frac{1}{\pi W_i^2}\int\dif\phi\int(r/W_i)^{2n}\exp(-r^2/W_i^2) r\dif r = \int z^n \exp(-z) \dif z = \Gamma(n+1).
\end{equation}

\begin{figure}[htpb]
    \centering
    \includegraphics[width=0.7\textwidth]{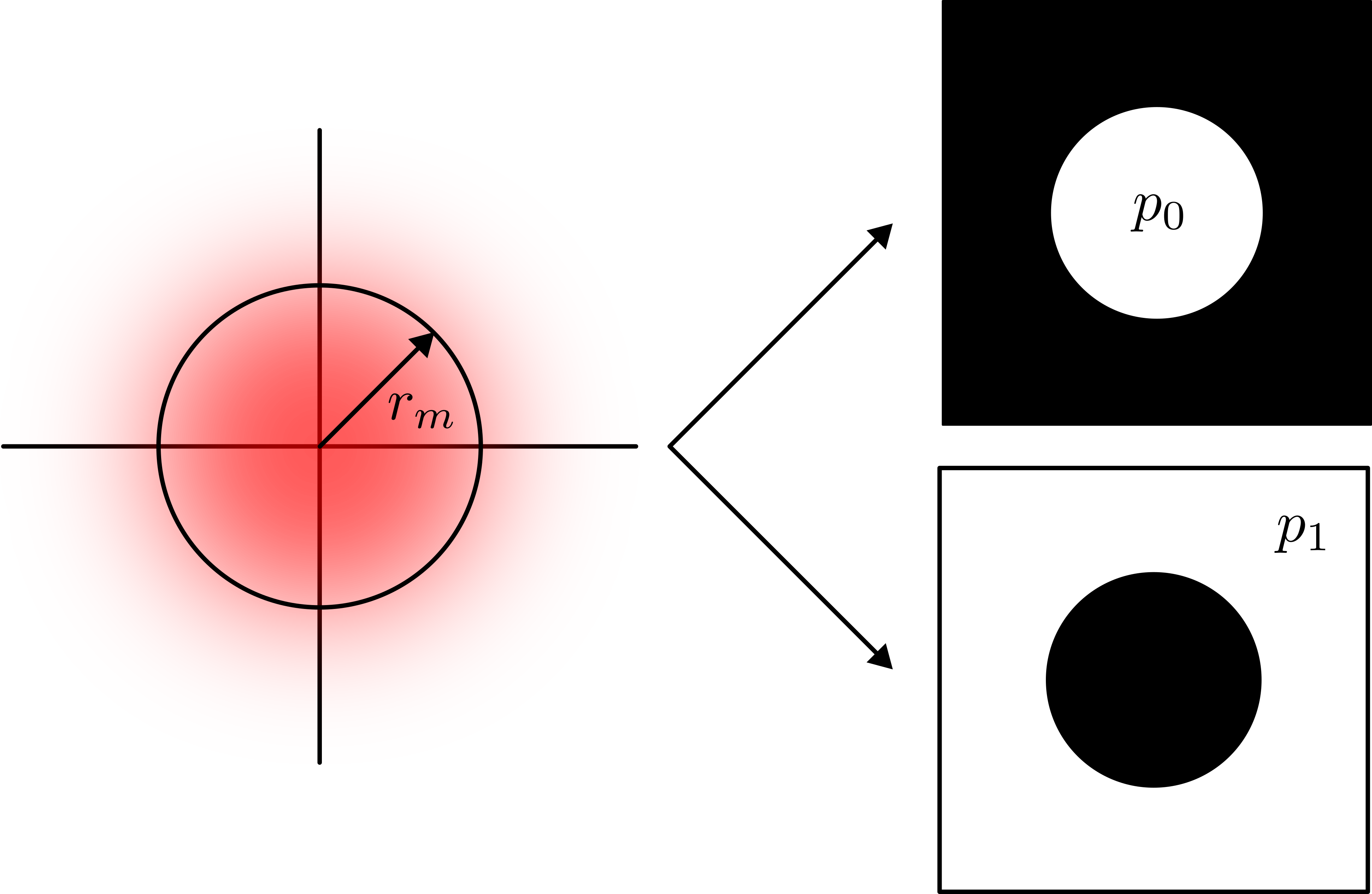}
    \caption{Left: Mask function of radius $r_m$ separates image into an inner
    and outer bucket detector. Right: Two detective areas represented by $p_0$ and
    $p_1$.}
    \label{fig:GaussianSpot}
\end{figure}

In general it will be impractical to have a high resolution single photon detector. So we consider a detection scheme that breaks up the image plane as shown in \autoref{fig:GaussianSpot} into two regions parameterized by a radius $r_m$.   The inner region is defined by the region of the image plane with radial coordinate $r < r_m$. The probability that a photon lands in this region is given by
\begin{equation}
p_0 = \frac{1}{\pi W_i^2}\int_0^{2\pi}d\phi\int_0^{r_m} \Braket{I(\mathbf r)}r\dif r
    = \int_{-(r_m/W_i)^2}^0 e^z \dif z
    = 1 - \exp(-r_m^2/W_i^2).
\label{eqn:2DGaussianErf}
\end{equation}
Likewise the outer region is the area $r>r_m$ and has a probability $p_1 = 1 - p_0 = \exp(-r_m^2/W_i^2)$.

\begin{figure}[htpb]
    \centering
    \includegraphics[width=0.8\textwidth]{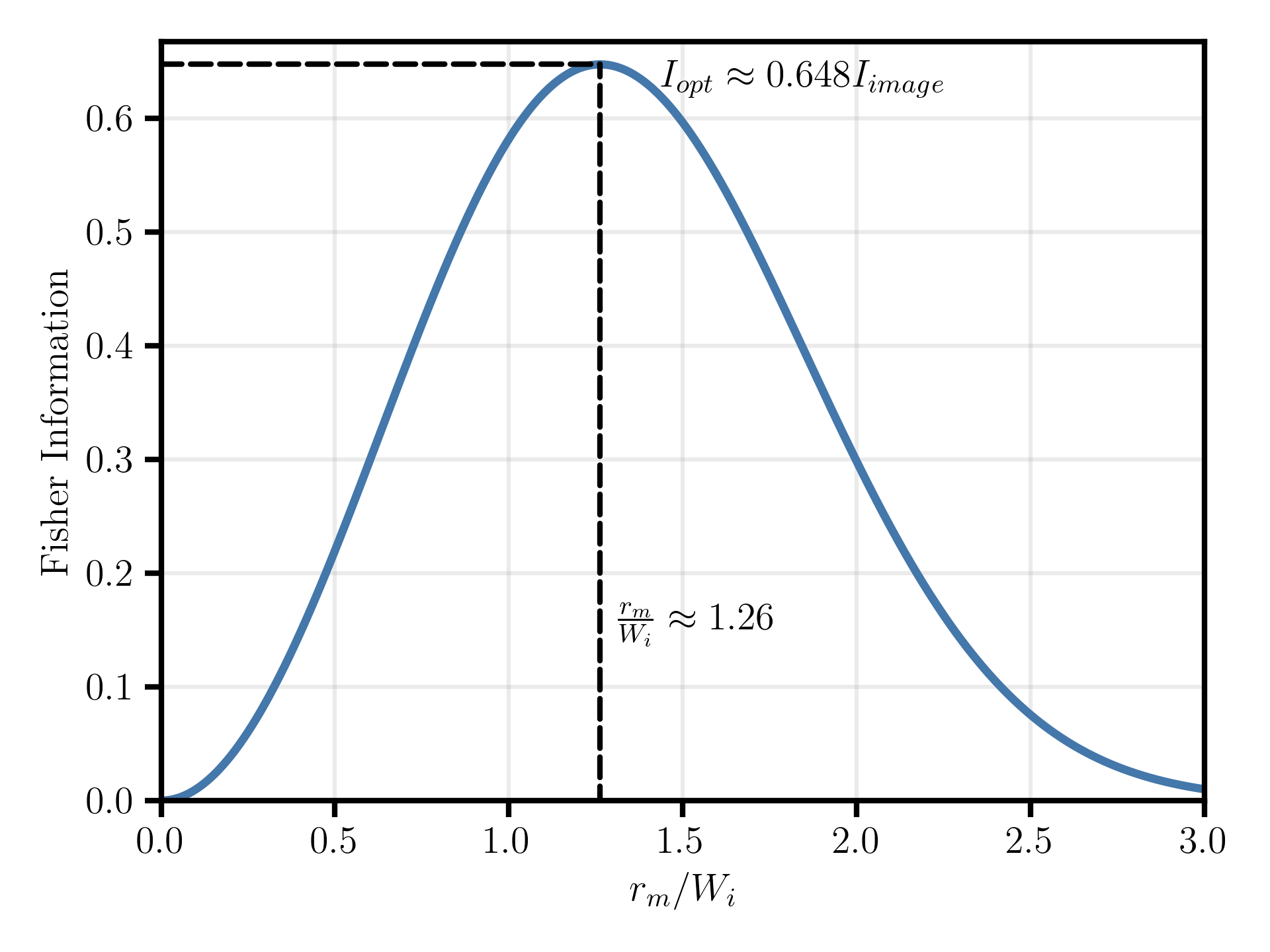}
    \caption{Maximum of $I(\theta)$ shows an ideal mask size of $r_m/W_i
             \approx 1.26$.}
    \label{fig:IdealMaskSize}
\end{figure}

Now, the Fisher Information in this two detector scheme is
\begin{equation}
\begin{split}
\mathcal I_2(\theta)&= \sum_{n=0,1} \left(\dpd{}{\theta}\log p_n\right)^2p_n
     = \frac{r_m^4}{W_i^4}\frac{\exp(-r_m^2/W_i^2)}{1-\exp(-r_m^2/W_i^2)}\mathcal I_\text{image},
\end{split}
\end{equation}
where $\mathcal I_\text{image}$ is the Fisher Information given in the full field image given in \autoref{eq:I_image}. The optimal mask size $r_m$ can be found, and is given by $r_m/W_i \approx 1.2624$ as shown in \autoref{fig:IdealMaskSize}. Note, the optimal mask has roughly 0.6476 the information content of the full image. Therefore, we only need two detectors to capture the majority of the information contained within the image plane.

\subsection{Signal Model}\label{sec:signal_model}
The above sections show how turbulence modifies the shape of the outgoing and returning lidar pulse and how recording the spatial distribution of arriving photons contains information regarding the refractive index structure constant. In order to quantify mechanical turbulence by sensing beam distortion, we need to relate the refractive index structure constant to the eddy dissipation rate (EDR), which is the primary measure used to quantify the turbulence experienced by aircraft \cite{FAA-H-8083-28A}. The eddy dissipation rate is related to the energy dissipation rate $\epsilon = \rm{EDR}^3$, which is the quantity generally used by the fluid dynamics community to quantify turbulence intensity. EDR has units m$^{2/3}$~s$^{-1}$ and $\epsilon$ has units m$^{2}$~s$^{-3}$. The remainder of the derivation linking energy dissipation rate $\epsilon$ with the refractive index structure parameter $C_n^2$ largely follows \cite{jumper}. 

$C_n^2$ is related to the structure constant for temperature fluctuations $C_T^2$ by
\begin{equation}\label{eq:CN2_CT2}
    C_n^2 \approx \left(7.9 \times 10^{-5} \frac{P}{\bar{T}^2}\right)^2 C_T^2,
\end{equation}
where $P$ and $\bar{T}$ are the local mean pressure and absolute temperature in hPa and Kelvin (K) respectively. The temperature structure constant is related to the eddy dissipation rate by
\begin{equation}
    \epsilon = \left[\frac{\gamma C_T^2 g\theta}{\frac{d\theta}{\dif z}\bar{T}^2}\right]^{3/2},
\end{equation}
where $\theta$ is the potential temperature.  Defining the Brunt-V\"ais\"al\"a frequency $N = g \frac{d\theta}{\dif z} \theta^{-1}$, rearranging and combining with \autoref{eq:CN2_CT2} yields
\begin{equation}\label{eq:CN2_edr}
    C_n^2(\epsilon) \approx \Gamma \left( 7.9\times 10^{-5}\frac{P}{\bar{T}^2}\right)^2\left(\frac{\bar{T}}{g}\right)^2 \epsilon^{2/3}N^2.
\end{equation}
The parameter $\Gamma = \gamma^{-1}$ is the mixing efficiency for which we assume the well-established value $\Gamma = 0.2$ \cite{Peltier2003}. The Brunt-V\"ais\"al\"a frequency does vary depending on the flow regime. For the purposes of detecting CAT we assume the value $N = 0.02\ \textrm{rad/s}$ for stably stratified flows in the stratosphere \cite{Feneyrou2009}. It is worth noting that these relations all assume turbulence with a Kolmogorov-like spectrum appearing in stably stratified flows which is typical of CAT. This model is not valid in the case of an unstable atmosphere with $N^2 < 0$ (convection). We hope to investigate those cases in a future work.

We now describe the nominal detection scenario. The objective of this work is to assess the feasibility of remote CAT detection by proxy measurement of EDR via measurement of $C_n^2$. To that end, we assume that there is a turbulent layer of air that begins at some distant range $R_{turb}$ and extends infinitely (or effectively to the maximum range of the lidar). Prior to this layer, the turbulence is assumed to be negligible. If we emit a lidar pulse and image the Rayleigh return from some point beyond the turbulence onset, $R_{image}$ onto a detector plane then the received signal energy is given by $E_{signal}$ according to \autoref{eq:total-returned-energy},
\begin{equation}
    E_{signal} \approx \beta P_0 \tau \Psi(R_{image}(t)) 10^{-2 \alpha R_{image}(t)} \Delta l,
\end{equation}
where $\beta$ is the scattering coefficient, $P_0$ is the peak pulse power, $\tau$ is the pulse duration, $\Psi(R_{image})$ is the integral of the Rayleigh phase function over the aperture solid angle, $\alpha$ is the attenuation coefficient, and $\Delta l$ is the length of the range cell. The range $R_{image}$ is described as a function of time because the craft is assumed to be approaching the turbulent layer with constant airspeed. The long-term average waist of the imaged return is given by $W_i$ following \autoref{eq:W_image}. We assume that the detector and associated optics are arranged so that all photons falling within a radius of $r_{in}$ at the image plane are counted by one detector and all photons falling within a radius $r_{in}< r < r_{out}$ are counted by a second detector. If the total per pulse returned energy is given by $E_{signal}$ then we can compute mean signal photon counts for the inner and outer detectors by converting $E_{signal}$ to an expected number of photons, 
\begin{equation}\label{eq:signal_photons}
\mu_{signal}(t) = E_{signal}(t) \frac{\lambda}{hc},
\end{equation}
and then dividing that mean count among the inner and outer detectors, following the probability relation in \autoref{eqn:2DGaussianErf},
\begin{equation}
\begin{aligned}
    p_{\rm in} = 1 - \exp\Big(-\frac{r_{\rm in}^2}{W_i^2}\Big), p_{\rm out} = 1 - \exp\Big(-\frac{r_{\rm out}^2}{W_i^2}\Big), \\
    \mu_{\rm in,signal} = \mu_{\rm signal} \, p_{\rm in}, \mu_{\rm out,signal} = \mu_{\rm signal} \, (p_{\rm out} - p_{\rm in}).
\end{aligned}
\end{equation}
For a single pulse, each detector experiences a total photon flux that is the sum of the signal term and an environmentally dependent background (BG) term,
\begin{equation}\label{eq:combined_rate}
    \mu_{in|out}(t) = \mu_{in|out, sig}(t) + \mu_{in|out,BG}.
\end{equation}
We assume for simplicity that $\mu_{*,BG}$ is constant in time though only minor modifications would be required to account for a time-varying background flux. Note also that in a real scenario there would be various efficiency factors that would reduce the expected count of recorded photons. We neglect these factors for this discussion but include estimates for them in the simulations that follow. 

The signal photon flux rates are dependent on EDR since as EDR increases, the relative proportion of signal photons that fall on the inner vs outer detectors will change following $W_i$. We propose a detection algorithm for EDR based on a simple Bayesian formulation,
\begin{equation}\label{bayes}
    \rho(\epsilon | N_{in},N_{out}) \propto \rho(N_{in},N_{out} | \epsilon) \rho(\epsilon),
\end{equation}
where $N_{in}$, $N_{out}$ are the counts observed by the inner and outer detectors. We assume that photon arrivals follow Poissonian statistics and therefore, for a single pulse, the likelihood term is given by,
\begin{equation}
    \rho(N_{in},N_{out}|\epsilon) = \frac{\mu_{in}^{N_{in}}\exp(-\mu_{in}) \mu_{out}^{N_{out}}\exp(-\mu_{out})}{N_{in}! N_{out}!}.
\end{equation}
Since the pulses are independent, the likelihood for multiple pulses is simply given by the product of the per-pulse likelihoods,
\begin{equation}\label{eq:likelihood}
\begin{aligned}
\rho(&N_{in|i=1...N_{pulses}},N_{out|i=1...N_{pulses}}|\epsilon) = \\
&\prod_{i=1}^{N_{pulses}} 
\frac{
\mu_{in}(t_i)^{N_{i,in}} \exp(-\mu_{in}(t_i)) \,
\mu_{out}(t_i)^{N_{i,out}} \exp(-\mu_{out}(t_i))
}{
N_{i,in}! \, N_{i,out}!
}.
\end{aligned}
\end{equation}

\begin{figure}[htpb]
    \centering
    \includegraphics[width=0.85\textwidth]{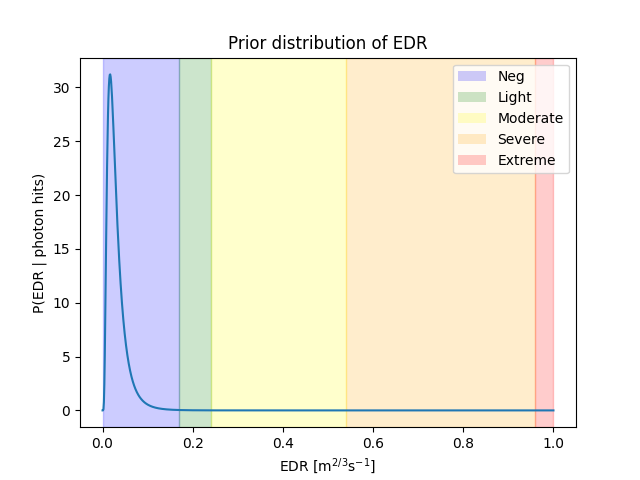}
    \caption{Prior distribution for $EDR = \epsilon^{1/3}$ with categories of turbulence for heavy aircraft indicated.}
    \label{fig:EDR_prior}
\end{figure}

For the prior we follow the survey in \cite{chang2023} and assume that the EDR is log-normally distributed,
\begin{equation}\label{prior}
\log(\mathrm{EDR}) = \log(\epsilon^{1/3}) \sim \mathcal{N}(-3.7 [\textrm{m}^{2/3}\textrm{s}^{-1}],0.6 [\textrm{m}^{4/3}\textrm{s}^{-2}]).
\end{equation}
Thus given a sequence of observations from each detector we can calculate a probability density function associated with EDR by \autoref{bayes} and \autoref{eq:likelihood}. By combining \autoref{bayes}, \autoref{eq:likelihood}, and \autoref{prior} we can integrate the received counts associated with a train of pulses into a posterior probability distribution for the EDR. Given sufficient photon counting statistics, the posterior distribution should peak at the true EDR provided sufficient evidence is accumulated to overcome the moderating influence of the prior. A plot of the prior distribution is shown in \autoref{fig:EDR_prior}.

\begin{table}[h!]
\centering
\begin{tabular}{lcccc}
\hline
\textbf{Aircraft Weight Class} & \textbf{Light} & \textbf{Moderate} & \textbf{Severe} & \textbf{Extreme} \\
(Max Takeoff Weight) & & & & \\
\hline
Light (< 15,500 lbs) & 0.13 & 0.16 & 0.36 & 0.64 \\
Medium (15,500 – 300,000 lbs) & 0.15 & 0.20 & 0.44 & 0.79 \\
Heavy (> 300,000 lbs) & 0.17 & 0.24 & 0.54 & 0.96 \\
\hline
\end{tabular}
\caption{Estimated $EDR=\epsilon^{1/3}$ thresholds for different aircraft weight classes \cite{aviationweather_gfa_help}.}
\label{tab:edr_thresholds}
\end{table}

\section{Experimental LiDAR Optical Efficiency Validation}

\subsection{Experiment}

\begin{figure}[h!]
    \centering
\includegraphics[width=\textwidth]{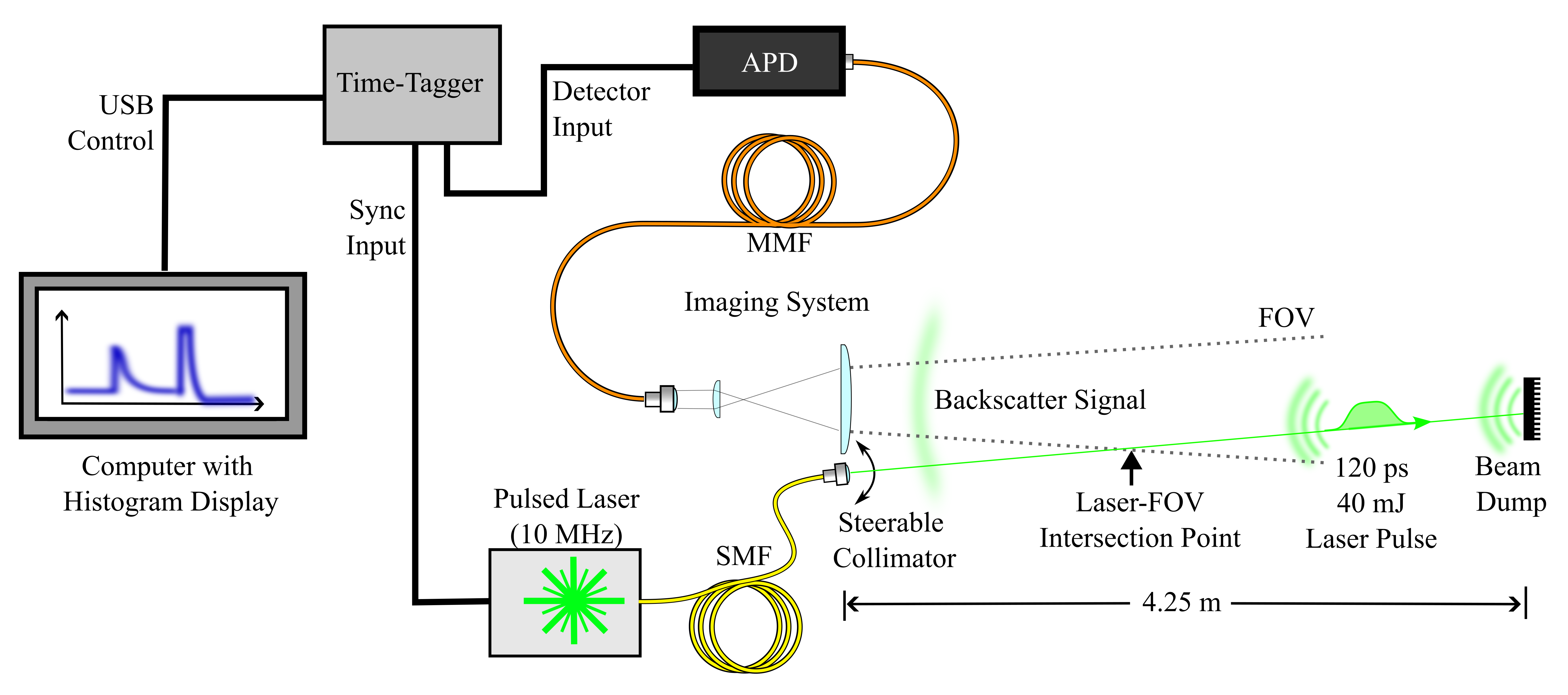}
\caption{One-hundred-twenty-picosecond pulses consisting of 40 mJ at 520 nm are generated with a Thorlabs picosecond gain-switch laser system at a 10 MHz rate. A steerable launching optic at the end of a single-mode fiber (SMF) varies the intersection point of the pulse at various distances with the collecting optic's field of view (FOV). A 3-inch (76.2 mm) collecting objective helps funnel scattered light into a multi-mode fiber (MMF). Detection events are registered as voltage pulses from an avalanche photodiode (APD) and are synchronized against electronic pulses emitted by the pulsed laser using a Hydraharp time-tagger. After integrating for 300 seconds, histograms are displayed on a computer. Multiple histograms were generated at various intersection distances. }
    \label{fig:exp_diagram}
\end{figure}

To validate our Rayleigh-based return link-budget model, we performed laboratory measurement using the experimental apparatus shown in Fig. \ref{fig:exp_diagram}. In that experiment, a pulsed laser (Thorlabs model GSL52A) emitted 520 nm pulses having a spectral linewidth of < 6 nm. The pulses were approximately 120 ps in temporal length, contained 40 mJ of energy, and were emitted at a repetition rate of 10 MHz from a single-mode fiber (SMF). Emitted pulses scattered from air molecules while a $\varnothing$76.2 mm objective lens collected and helped couple light into a $\varnothing$100 $\mu$m core multi-mode fiber (MMF). This MMF coupled light into a silicon-based Perkin Elmer avalanche photodiode (APD). Detector ``clicks" were collected by a Hydraharp time-tagger and synchronized against synchronization pulses emitted by the pulsed laser system. After integrating for 300 seconds, histograms were generated.

The operational range from the objective lens to the beam-dump was measured at 4.25 m. Because our APDs have a detector dead-time of $\approx$25 ns, any signal return is heavily biased near the laser's intersection distance to the collection optic's field of view (FOV). A laser pulse would need to travel an additional 7.5 meters before scattered photons could once again be registered after the detector resets for the next detection event. Naturally, this biases the expected return signal data. To overcome this dead-time biasing, the laser's launching collimator was placed on a steerable mount to alter the intersection range of the emitted pulses with the collection's FOV. This bi-static LiDAR differs from the mono-static LiDAR theory previously shown -- namely because the launching optic now has an offset from the center of the collecting optic. However, the two models converge in the limit of long-range scattering.

\begin{figure}[h!]
    \centering
\includegraphics[width=\textwidth]{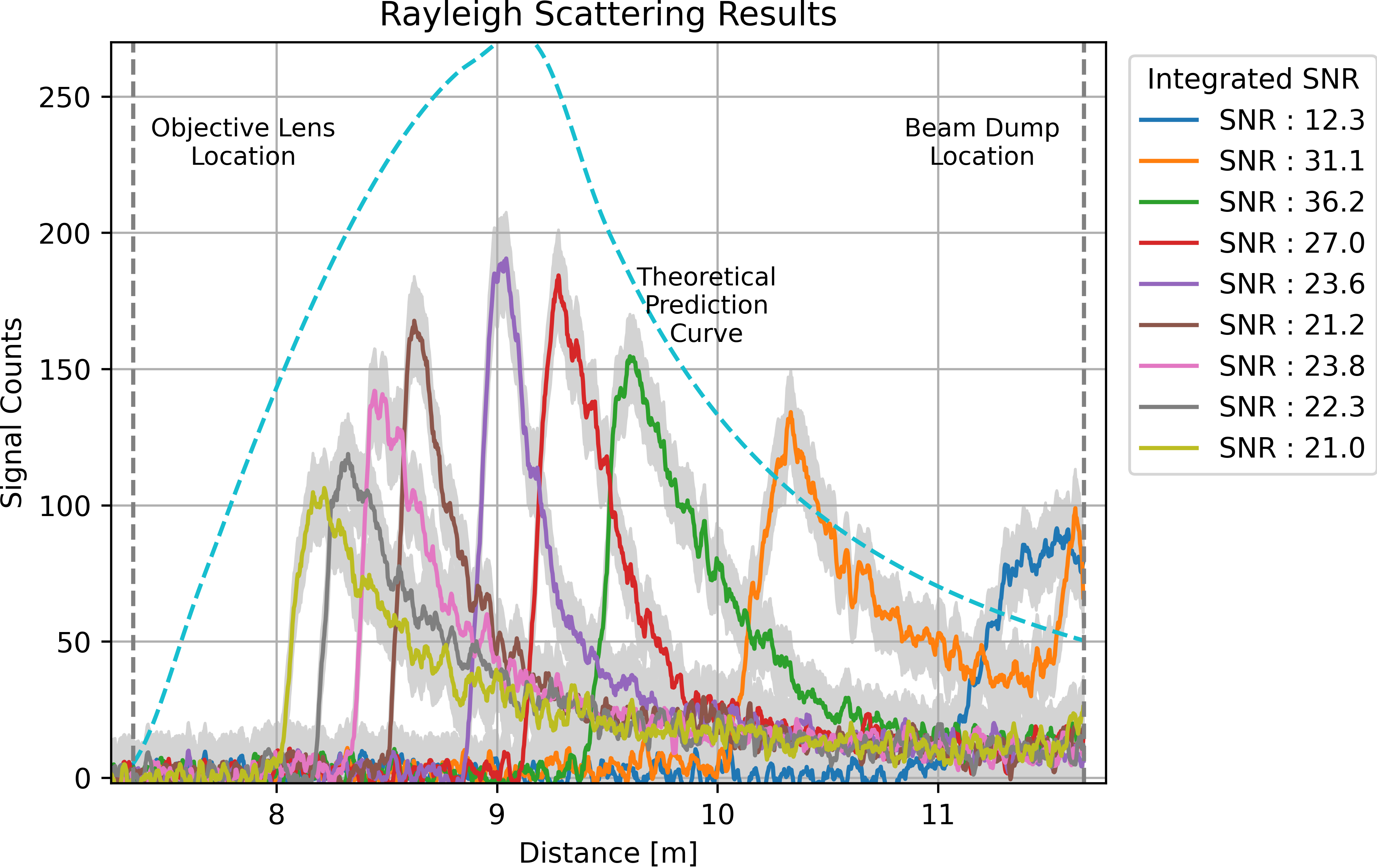}
\caption{Multiple histograms were generated with varying intersection distances to the collection system's field of view (solid lines). An ``expected-counts" curve using our theoretical optical efficiency model and experimental parameters is also shown (dashed curve). Light gray curves designate Poisson noise around each solid curve. After integrating over all elevated time-bins above background, an effective signal-to-noise (SNR) for each histogram is generated when assuming Poisson noise. }
    \label{fig:data}
\end{figure}

Various histograms were generated at various distances and the resulting envelope shows a reasonable fit to the theoretical model. Because there exist multiple variables that can be tweaked (e.g., the fiber coupling efficiency, the APD detection efficiency, and the launching offset distance from the objective-lens center), the model can be overfit. Nonetheless, the general trend in the far field is of primary interest and shows reasonable correlation with the distant histograms. Data histograms and a theoretical optical efficiency curve are shown in \autoref{fig:data}.      

\subsection*{Analysis of Air Scattering in a Short-Range Single-Photon LiDAR Experiment}

From \autoref{fig:data}, we can estimate the detector's effective FOV using the $1/e^2$ peak intensity value of the longest pulse. This effective FOV is essentially generated by the detector's long dead time and the path-dependent collection efficiency that effectively biases the detection of photons to regions closer to the detector within the entire FOV. Using the detector count data, that value is estimated to be $0.73 \textrm{m}$. At any given time, our detector can only detect backscattered photons from $0.73 \textrm{m}$ of the entire $4.25 \textrm{m}$ path length. Swiveling the steerable collimator in \autoref{fig:exp_diagram} allows us to probe various scattering regions along the $4.25 \textrm{m}$ optical path. Using this $0.73 \textrm{m}$ detection length, we can estimate the beam illumination volume to compare scattering theory with our measured results.  


Since the pulse period (\SI{100}{ns}) exceeded the detector dead time (25 ns), each pulse 
constituted an independent scattering trial. Over \SI{300}{s} at 
\SI{10}{MHz}, the total number of emitted pulses was

\begin{equation}
N_{\text{pulses}} = (10^7 \,\text{s}^{-1})(300 \,\text{s})
= 3.0 \times 10^9.
\end{equation}
After background subtraction, the strongest integrated return signal (comprised of 27{,}544 detection events) was obtained by integrating within the 
scattering peak -- corresponding to the illuminated \SI{0.73}{m} segment. 
The per-pulse detection probability is therefore

\begin{equation}
P_{\mathrm{det}} =
\frac{27{,}544}{3.0 \times 10^9}
= 9.18 \times 10^{-6}.
\end{equation}
The illuminated beam volume was

\begin{equation}
V = \pi r^2 L
= \pi (1.05\times10^{-3}\,\text{m})^2 (0.73\,\text{m})
\approx 2.53\times10^{-6}\,\text{m}^3.
\end{equation}
These quantities are then used to compared our measured results with both Rayleigh and Mie scattering.  

\subsubsection*{Rayleigh Scattering Prediction}

The Rayleigh scattering cross section per air molecule at visible wavelengths 
is well established and scales as $\lambda^{-4}$ \cite{Bohren}. 
Evaluating the standard expression for air at \SI{520}{nm} gives

\begin{equation}
\sigma_R \approx 5 \times 10^{-31} \,\text{m}^2.
\end{equation}
Using a molecular number density of air under laboratory conditions,

\begin{equation}
n_{\text{air}} \approx 2.5 \times 10^{25} \,\text{m}^{-3},
\end{equation}
the Rayleigh volume scattering coefficient becomes

\begin{equation}
\beta_R = n_{\text{air}} \sigma_R
\approx 1.25 \times 10^{-5} \,\text{m}^{-1}.
\end{equation}
The optical depth over the observed \SI{0.73}{m} segment is

\begin{equation}
\tau = \beta_R L
= (1.25 \times 10^{-5})(0.73)
\approx 9.13 \times 10^{-6}.
\end{equation}
For very small optical depths ($\tau \ll 1$), the Beer--Lambert law gives

\begin{equation}
P_{\text{scatter}} = 1 - e^{-\tau} \approx \tau,
\end{equation}
which is the standard first-order approximation used in radiative transfer theory \cite{Chandrasekhar}. Thus, the predicted probability that a photon scatters within the observed segment is

\begin{equation}
P_{\text{scatter,total}}
\approx 9.13 \times 10^{-6}.
\end{equation}

\subsubsection*{Comparison with Measurement}

The experimentally measured per-pulse detection probability is

\begin{equation}
P_{\mathrm{det}} = 9.18 \times 10^{-6},
\end{equation}
which is essentially identical to the predicted Rayleigh optical depth 
over the same path length.

The detected probability is more generally written as

\begin{equation}
P_{\mathrm{det}}
= \beta_R L
\left(\frac{\Delta\Omega}{4\pi}\right)
\eta,
\end{equation}
where $\Delta\Omega$ is the collection solid angle and $\eta$ is the total 
optical and detector efficiency. The use of a $\varnothing$76.2 mm collecting objective lens a few meters from FOV intersection point with the beam line naturally implies $\Delta\Omega/4\pi < 1/2 $, so the measured detection probability is slightly larger than what would be explained by Rayleigh scattering alone. Nonetheless, the close numerical agreement between 
$P_{\mathrm{det}}$ and $\beta_R L$ demonstrates that molecular Rayleigh 
scattering alone is the dominant signal source and is enough to explain the observed signal magnitude.

\subsubsection*{Role of Mie Scattering}

Mie scattering from aerosol particles (dust) could contribute additional signal, 
since micron-scale particles possess much larger per-particle scattering cross 
sections than air molecules. However, the measured probability is already 
consistent with the full molecular Rayleigh optical depth. Therefore, while 
Mie scattering may provide a modest additive contribution, it is not required 
to account for the magnitude of the observed signal.

\subsubsection*{Conclusion}

The measured single-pulse detection probability,
\[
P_{\mathrm{det}} \sim 9 \times 10^{-6},
\]
matches the predicted Rayleigh optical depth of air over 
\SI{0.73}{m} at \SI{520}{nm}. This strong quantitative agreement indicates 
that the observed LiDAR peak is well explained by molecular Rayleigh scattering, 
with any contribution from aerosol (Mie) scattering likely representing only a 
secondary perturbation.

\section{Simulation Results}
\begin{figure}[h!]
    \centering
%
%
%
%
%
%
%
%
%
%
%

\includegraphics[width=\textwidth]{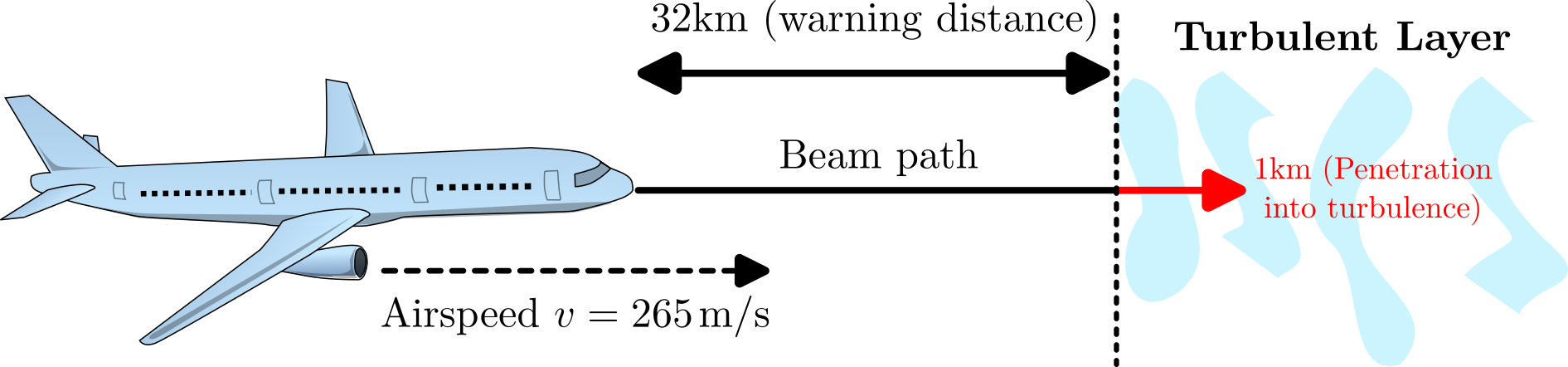}
\caption{Schematic of the aircraft approaching a turbulent layer. The beam path intersects the turbulent layer, and measurements are taken at a specific point inside.}
    \label{fig:scenario_geometry}
\end{figure}

We now apply the proposed detection theory to a simulated scenario with plausible first order hardware parameters. The parameters for the simulation are all specified in \autoref{sec:sim_params}. We assume as in \autoref{sec:signal_model} that the aircraft is approaching a layer of increased turbulence with an airspeed of $265~\mathrm{m/s}$ representing typical cruising speed. We specify a desired warning time of 2 minutes. That is, the accumulated photon statistics must be sufficient to detect the turbulent layer when it is still $\sim 32~\mathrm{km}$ from the aircraft. We also set a fixed integration time of $40$ seconds. Additionally, since the method relies on measuring beam spread, the beam must be measured at some point within the turbulent parcel. The penetration distance into the layer is a tunable parameter in the simulation code that we set to $1~\mathrm{km}$. A schematic of the scenario setup is shown in \autoref{fig:scenario_geometry}. The inner and outer detector regions are parameterized by their respective radii. The inner detector radius $r_{inner}$ was chosen to correspond to the anticipated beam spread for the case in which there is no turbulence present from \autoref{eq:W_image}. The outer detector radius was chosen as $r_{outer} = 2r_{inner}$. 

\begin{figure}[h!]
 
\includegraphics[width=\textwidth]{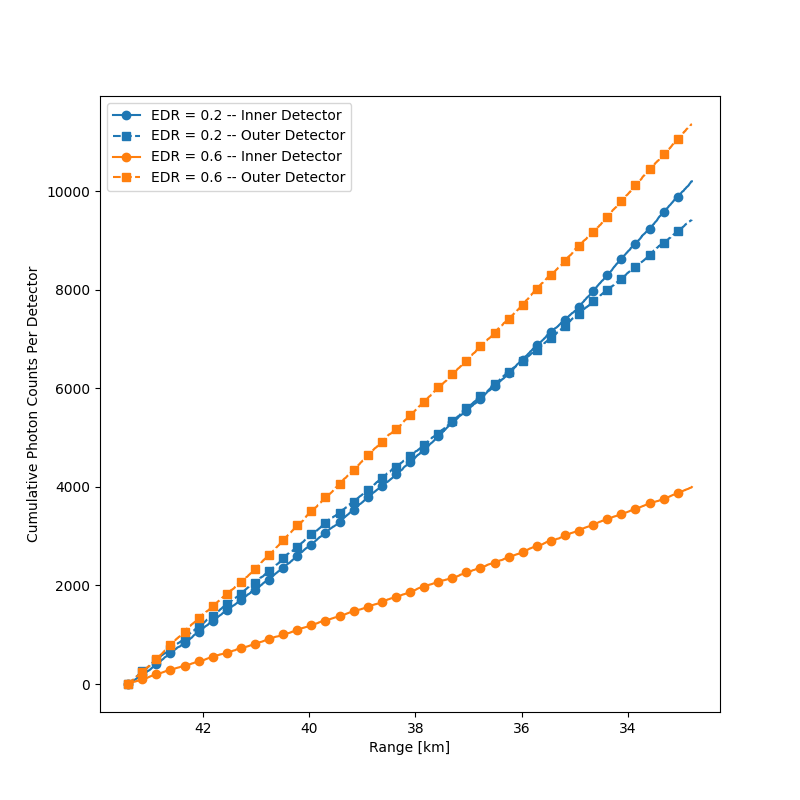}
\caption{Simulated accumulated counts in each detector region as a function of range and turbulence level}
    \label{fig:signal_vs_range.png}
\end{figure}

For each pulse as the aircraft approaches the layer, we compute the image plane long term beam width $W_i$ from \autoref{eq:W_image} as well as the signal energy from \autoref{eq:total-returned-energy} and the background energy. Representative values for $\bar{T}$ and $P$ were found for a $9$ km altitude via MODTRAN with the standard atmosphere profile. The local air density is computed from these two via the ideal gas law. The scattering cross section of air was taken from \cite{sneep2005direct} with the volumetric scattering coefficient $\beta$ computed from this value and the density $\rho$. The attenuation coefficient, $\alpha$, and background radiance terms $L_{bg}$ were also computed using the MODTRAN standard atmosphere. The background photon flux rate noise term is computed by integrating the MODTRAN radiance value $L(x,\omega) = L_{bg}$ over the field of view of each detector and a spectral window, $BW_{detector}$ corresponding to the passband of a narrow filter,
\begin{equation}\label{eq:background_rate}
\mu_{inner|outer,BG} = \tau A\frac{\lambda}{hc} BW_{detector}L_{bg}\int_{FOV[inner|outer]}\cos(\theta) \dif\Omega,
\end{equation}
where $A$ is the area of the primary aperture. In simulation we compute a single background flux for the entire detector setup, i.e., both inner and outer detectors and then divide the total background radiance according to the relative areas of the inner and outer detectors. The background radiance is derived from MODTRAN using the standard atmosphere profile. The MODTRAN simulation was configured with the sun positioned with a $60^\circ$ elevation and zero azimuth with respect to the sensor's line of sight. These parameters were selected to represent typical daytime operation where the sun is overhead but not directly in the line of sight of the sensor. The final combined signal plus noise photon flux rates for each detector are computed by \autoref{eq:combined_rate}, with the modification that we apply a constant efficiency factor to the rates of both detectors to account for losses in the optical system. For the simulations below, we take this efficiency factor to be $T = 0.08$. Finally, using the photon flux terms we simulate photon counts in the inner and outer detector following Poisson statistics over the range gate integration time $\delta z / c$. The basic mode of detection is shown by the plot in \autoref{fig:signal_vs_range.png}. As the aircraft approaches the turbulent pocket, turbulence within the pocket causes the beam to spread and an excess of counts to be recorded by the outer detector relative to the inner detector.

\begin{figure}[htpb]
    \centering
    \includegraphics[width=0.85\textwidth]{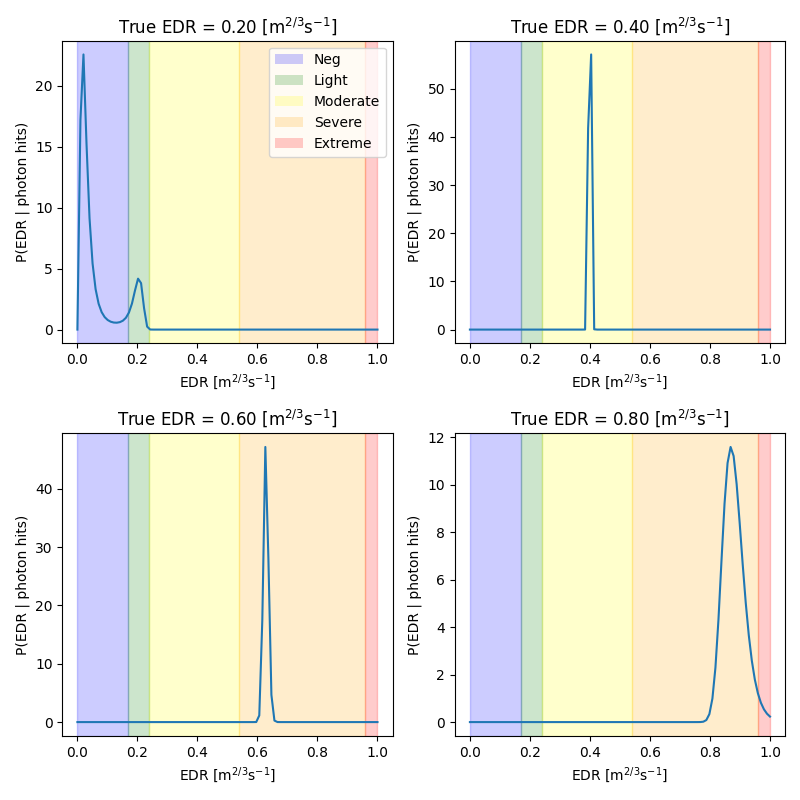}
    \caption{Computed posterior distributions for EDR derived from simulated photon counts.}
    \label{fig:EDR_posterior}
\end{figure}

Using the described simulation setup, we set a value for the EDR and simulate pulse-to-pulse counts on the inner and outer detector regions and the aircraft approaches the turbulent layer. We cap the number of received photons at one per pulse per detector to simulate the effect of detector dead time. This effect is not accounted for in the detector theory above but it is a rare enough occurrence that it does not distort the statistics. Each pulse contributes to the likelihood term following  \autoref{eq:likelihood} and at the end of the simulation, when the aircraft is at the required warning distance, we output the posterior accumulated via \autoref{bayes}. Computed posterior distributions for various values of EDR are shown in \autoref{fig:EDR_posterior}. There is some difficulty in detecting very small levels of turbulence and difficulty precisely characterizing the level of severe turbulence. However, as a general statement, the proposed method is able to detect elevated turbulence at the target range with high confidence and characterize the EDR value with good precision. 

\section{Conclusion}
We have introduced a differential detection system using Rayleigh lidar that estimates turbulence strength by measuring beam broadening at range and mapping the observed image-plane counts to a posterior distribution for the eddy dissipation rate (EDR). Unlike Doppler lidar, the approach does not rely on aerosols; it remains effective at cruise altitudes; unlike prior Rayleigh lidar methods that measure small intensity variations (on top of the baseline return) from density-fluctuations, our approach exploits spatial broadening of the beam, moving signal photons outside the nominal beam, improving shot-noise performance.  Under modest assumptions and size, weight, and power (SWAP) constraints, our analysis and simulations indicate detection of moderate CAT at ranges exceeding 30~km, providing roughly two minutes of warning time.

\FloatBarrier

\begin{backmatter}
\bmsection{Funding}
Funding for this work was provided by MITRE's Independent Research and Development Program.

\bmsection{Disclosures}
The authors declare no conflicts of interest.

\bmsection{Data availability} Data underlying the results presented in this paper are not publicly available at this time but may be obtained from the authors upon reasonable request.

\end{backmatter}

\bibliography{References}

\begin{thebibliography}{10}
\newcommand{\enquote}[1]{``#1''}

\bibitem{sharman2012}
R.~D. Sharman, S.~B. Trier, T.~P. Lane, and J.~D. Doyle, \enquote{Sources and
  dynamics of turbulence in the upper troposphere and lower stratosphere: A
  review,} {\protect\JournalTitle{Geophysical Research Letters}} \textbf{39}
  (2012).

\bibitem{lee2019}
S.~H. Lee, P.~D. Williams, and T.~H.~A. Frame, \enquote{Increased shear in the
  north atlantic upper-level jet stream over the past four decades,}
  {\protect\JournalTitle{Nature}} \textbf{572}, 639--642 (2019).

\bibitem{prosser2023}
M.~C. Prosser, P.~D. Williams, G.~J. Marlton, and R.~G. Harrison,
  \enquote{Evidence for large increases in clear-air turbulence over the past
  four decades,} {\protect\JournalTitle{Geophysical Research Letters}}
  \textbf{50}, e2023GL103814 (2023). E2023GL103814 2023GL103814.

\bibitem{smith-williams}
I.~Smith, P.~Williams, and R.~Schiemann, \enquote{Clear-air turbulence trends
  over the {North Atlantic} in high-resolution climate models,}
  {\protect\JournalTitle{Climate Dynamics}} \textbf{61}, 1--17 (2023).

\bibitem{zhao2023a}
J.~Zhao, X.~Luo, Z.~Liu, \emph{et~al.}, \enquote{{Influence of airborne LiDAR
  wavelength on the detection distance of clear air turbulence},}
  {\protect\JournalTitle{Optical Engineering}} \textbf{62}, 114101 (2023).

\bibitem{matayoshi2018}
N.~Matayoshi \emph{et~al.}, \enquote{{Development and flight demonstration of a
  new lidar-based onboard turbulence information system},}  (2018). {31st
  Congress of the International Council of the Aeronautical Sciences, Brazil}.

\bibitem{vrancken2016}
P.~Vrancken, M.~Wirth, G.~Ehret, \emph{et~al.}, \enquote{Airborne
  forward-pointing {UV Rayleigh} lidar for remote clear air turbulence
  detection: system design and performance,} {\protect\JournalTitle{Appl.
  Opt.}} \textbf{55}, 9314--9328 (2016).

\bibitem{zhao2023b}
J.~Zhao, X.~Luo, and H.~Liu, \enquote{An airborne visible light lidar’s
  methodology for clear air turbulence detection based on weak optical signal,}
  {\protect\JournalTitle{Photonics}} \textbf{10} (2023).

\bibitem{sneep2005direct}
M.~Sneep and W.~Ubachs, \enquote{Direct measurement of the {Rayleigh}
  scattering cross section in various gases,} {\protect\JournalTitle{Journal of
  Quantitative Spectroscopy and Radiative Transfer}} \textbf{92}, 293--310
  (2005).

\bibitem{Andrews2023}
L.~C. Andrews and M.~K. Beason, \emph{Laser beam propagation in random media:
  new and advanced topics} (2023).

\bibitem{Rodenburg2014}
B.~Rodenburg, M.~Mirhosseini, M.~Malik, \emph{et~al.}, \enquote{Simulating
  thick atmospheric turbulence in the lab with application to orbital angular
  momentum communication,} {\protect\JournalTitle{New Journal of Physics}}
  \textbf{16}, 033020 (2014).

\bibitem{Rodenburg2015}
B.~Rodenburg, \enquote{Communicating with transverse modes of light,} Ph.D.
  thesis, University of Rochester (2015).

\bibitem{TurbulenceOTF2017}
R.~C. Hardie, J.~D. Power, D.~A. LeMaster, \emph{et~al.}, \enquote{Simulation
  of anisoplanatic imaging through optical turbulence using numerical wave
  propagation with new validation analysis,} {\protect\JournalTitle{Optical
  Engineering}} \textbf{56}, 071502--071502 (2017).

\bibitem{FAA-H-8083-28A}
{Federal Aviation Administration}, \emph{Aviation Weather Handbook
  (FAA‑H‑8083‑28A)} (Federal Aviation Administration, U.S. Department of
  Transportation, Washington, DC, USA, 2024), revised december 12, 2024 ed.
  Supersedes FAA‑H‑8083‑28 (2022).

\bibitem{jumper}
G.~Jumper, J.~Roadcap, and P.~Tracy, \enquote{Estimating velocity turbulence
  magnitudes using the thermosonde,} {\protect\JournalTitle{AIAA, 41st
  Aerospace Sciences Meeting, Reno, Nevada}} p.~12 (2003).

\bibitem{Peltier2003}
W.~R. Peltier and C.~P. Caulfield, \enquote{Mixing efficiency in stratified
  shear flows,} {\protect\JournalTitle{Annual Review of Fluid Mechanics}}
  \textbf{35}, 135--167 (2003).

\bibitem{Feneyrou2009}
P.~Feneyrou, J.-C. Lehureau, and H.~Barny, \enquote{Performance evaluation for
  long-range turbulence-detection using ultraviolet lidar,}
  {\protect\JournalTitle{Appl. Opt.}} \textbf{48}, 3750--3759 (2009).

\bibitem{chang2023}
H.-C. Ko, H.-Y. Chun, R.~Sharman, and J.-H. Kim, \enquote{Comparison of eddy
  dissipation rate estimated from operational radiosonde and commercial
  aircraft observations in the {United States},} {\protect\JournalTitle{Journal
  of Geophysical Research: Atmospheres}} \textbf{128} (2023).

\bibitem{aviationweather_gfa_help}
{Aviation Weather Center}, \enquote{{Graphical Forecasts for Aviation (GFA)
  Help — Products},} \url{https://aviationweather.gov/gfa/help/#products}
  (2025). Accessed: 2025-11-21.

\bibitem{Bohren}
C.~F. {Bohren} and D.~R. {Huffman}, \enquote{{Absorption and scattering of
  light by small particles},} Research supported by the University of Arizona
  and Institute of Occupational and Environmental Health. New York,
  Wiley-Interscience, 1983, 541 p. (1983).

\bibitem{Chandrasekhar}
S.~{Chandrasekhar}, \emph{{Radiative transfer}} (1960).

\end{thebibliography}

\appendix

\section{Simulation Parameters}\label{sec:sim_params}
\begin{table}[h!]
\centering
\caption{Simulation Parameters}
\begin{tabular}{llc}
\hline
\textbf{Category} & \textbf{Parameter} & \textbf{Value} \\
\hline
\multicolumn{3}{l}{\textbf{Environmental}} \\
$\bar{T}$ & Average air temperature at altitude & $225~\mathrm{K}$ \\
$P$ & Air pressure at altitude & $300~\mathrm{hPa}$ \\
$\rho_{\mathrm{air}}$ & Air density & $16.03~\mathrm{mol/m^3}$ \\
$\sigma$ & Scattering cross section of air & $5.1\times10^{-31}~\mathrm{m^2}$ \\
$I_{\mathrm{bg}}$ & Background radiance & $0.1~\mathrm{W/(m^2\,sr\,nm)}$ \\
$\alpha$ & Attenuation coefficient & $0.007~\mathrm{km^{-1}}$ \\[4pt]

\multicolumn{3}{l}{\textbf{System}} \\
$P_{avg}$ & Average transmitted power & $2.25~\mathrm{W}$ \\
$\lambda$ & Laser central wavelength & $532~\mathrm{nm}$ \\
$\mathrm{PRF}$ & Pulse repetition frequency & $3000 \textrm{Hz}$ \\
$w_0$ & Beam launch waist & $3.5~\mathrm{cm}$ \\
$\tau$ & Pulse duration & $3~\mathrm{ns}$ \\
$\delta z$ & Range gate depth & $50~\mathrm{m}$ \\
$\mathrm{BW}_{\mathrm{detector}}$ & Optical passband width & $1.2~\mathrm{nm}$ \\
$f$ & Optical focal length & $300~\mathrm{mm}$ \\
$D$ & Entrance aperture diameter & $200~\mathrm{mm}$ \\
$Mag$ & System magnification factor & $100x$\\
$T$ & Optical System Transmissivity & $0.08~\mathrm{unitless}$\\ 
$v$ & Airspeed / rate of closure with turbulent parcel & $265~\mathrm{m/s}$\\
$t_{int}$ & Integration Time & $40~\mathrm{s}$\\
\hline
\end{tabular}
\label{tab:sim_params}
\end{table}


\end{document}